\documentclass[epj]{svjour}
\usepackage{graphics}
\usepackage{bm}

\newcommand{\x}{{\bm x}}

\newcommand{\be}{\begin{equation}}
\newcommand{\ee}{\end{equation}}
\begin{document}
\title{Lattice Boltzmann simulations of capillary filling: finite vapour density effects}
\subtitle{Lattice boltzmann simulations of capillary filling}

\author{F. Diotallevi\inst{1} \and L.Biferale\inst{2} \and S. Chibbaro\inst{3} \and G. Pontrelli\inst{1} \and F. Toschi\inst{1} \and S. Succi\inst{1}}
\institute{Istituto per le Applicazioni del
Calcolo CNR, Viale del Policlinico 137, 00161 Roma, Italy. 
\and Dept. of Physics and INFN, University of Tor Vergata, Via della Ricerca Scientifica 1, 00133 Roma, Italy. 
\and Dept. of Mechanical Engineering, University of Tor Vergata, Viale
Politecnico 8, Rome, Italy. 
}
\date{Received: date / Revised version: date}
%
\abstract{Numerical simulations of two-dimensional
capillary filling using the pseudo-potential lattice Boltzmann model
for multiphase fluids are presented, with special emphasis on the role
of finite-vapour density effects. 
It is shown that whenever the density of the light-phase exceeds about 
ten percent of the dense phase, the front motion proceeds through a combined effect 
of capillary advection and condensation.
As a result, under these conditions, the front proceeds at a higher speed as compared to 
the Washburn prediction.
It is suggested that such an acceleration effect might be observed in experiments
performed sufficiently close to critical conditions.
\PACS{ {83.50.Rp}{},\and {68.03.Cd}{}
     } 
} 
\authorrunning
\titlerunning
\maketitle

\section{Introduction}
\label{intro}

In the recent years, increasing attention has been paid to the use of 
lattice Boltzmann (LB) techniques beyond the strictly hydrodynamic regime, i.e. for micro and nano-fluidic
applications. LB holds great potential for these
applications because it can offer an optimal compromise between the physical realism
of atomistic methods and the computational efficiency of continuum fluid mechanics.
However, turning this potential into a robust and reliable prediction tool for
complex micro/nanofluidic problems, requires a detailed and {\it quantitative} validation
program on a number of test and benchmark problems.
In this work, we present one such validation study, as applied to the important
problem of capillary filling.
Capillary filling is an old problem, originating with
the pioneering works of Washburn \cite{washburn} and Lucas
\cite{Lucas}. Recently, with the explosion of theoretical,
experimental and numerical works on microphysics and nanophysics, the
problem attracted considerable renewed interest
\cite{degennes,dussain,washburn_rec,tas}.  
Capillary filling is a typical ``moving contact line'' problem, in which
interface motion can only take place thanks to subtle 
non-hydrodynamic effects occurring at the contact point between 
liquid-gas and solid phase.
As a result, it offers an ideal testground for assessing the 
capabilities of mesoscopic methods, such as LB, to describe
phenomena beyond the hydrodynamic regime. 

\section{The Washburn description}

A simple and yet powerful picture of capillary motion was developed back in 20's
with the pioneering work of Washburn and Lucas.
The Washburn picture is easily derived by writing the momentum equation
for the mass of moving liquid in the capillary.
This leads to the following equation:
\begin{equation}
\rho_l \frac{d}{d \tilde t} (\tilde z \dot {\tilde z})  =  \frac{2 \gamma cos(\theta)}{H} - \frac{12 \mu_l \tilde z \dot {\tilde z}}{H^2}
\label{WASH1}
\end{equation}
where $\tilde z(t)$ is centerline position of the advancing front at time $\tilde t$, $H$ the transverse size
of the channel, $\gamma$ is the surface tension, $\theta$ is the static contact angle, $\rho_l$ the density 
and $\mu_l$ the  viscosity of the liquid.
The factor $12$ stems from the specific geometry considered here, i.e.
two infinite parallel plates separated by a distance $H$ -- see fig. \ref{fig:1}). 
More precisely, in the overdamped regime, the Washburn solution reads 
as follows:
\begin{equation}
\tilde z^2( \tilde t) - \tilde z^2(0) = \frac{\gamma H cos(\theta)}{3 \mu_l } \tilde  t 
\end{equation}
To the purpose of comparing with experimental results, it is useful to recast
this relation in dimensionless form (reduced units), $t=\tilde t/t_{cap}$ and $z = \tilde z /H$, being 
the capillary time $t_{cap} = H \mu/\gamma$.
This leads to the universal law:
\begin{equation}
\label{WASH}
z^2(t) - z^2(0) = \frac{cos(\theta)}{3} t.
\label{eq:washburn}
\end{equation}
The Washburn solution (\ref{WASH}) above spells troubles for microfluidic applications where
fast-fill is a goal.
Indeed, (\ref{WASH}) clearly shows that the front moves at increasingly lower speed as it propagates
down the capillary. The reason is easily understood by noting that, according
to the eq. \ref{WASH1}, the front position $z(t)$ evolves like 
the trajectory of a point-like particle, whose mass grows linearly with 
the position $z$, driven by a constant capillary 
force and damped by a dissipative force also growing linearly with $z$.
By neglecting inertia (overdamped regime), the capillary force competes with a linearly
increasing dissipation, which ultimately leads to a front speed $\dot z$ decaying like $1/z$. 
Strategies to do away with this basic limitation make the object of intense scientific and
technological research, such as interface functionalization, optimal coating, and related issues.
As already remarked in the literature \cite{washburn_rec}, the asymptotic
behaviour (\ref{eq:washburn}) hinges on a number of
simplifying assumptions, namely:
(i) the inertial terms in the Navier-Stokes equation are negligible,
(ii) the instantaneous {\it bulk} profile is given by the Poiseuille
flow, (iii) the microscopic slip mechanism which allows for the
interface motion is not relevant for bulk quantities (such as the
overall position of the interface inside the channel), (iv) inlet and
outlet phenomena can be neglected (limit of infinitely long channels);
(v) the liquid is filling in a capillary, either empty or filled with
gas whose total mass is negligible with respect to the liquid one. 
None of these assumptions needs to be true in actual experiments, and to
the purpose of comparison with experimental data, it is therefore important
to address these limitations within a suitably generalization of the
Washburn equation, to which we shall return shortly.
In this paper, we shall be interested specifically in the finite-vapour density issue (v).

\section{LBE for capillary filling}

The model used in this work is a suitable
adaptation of the Shan-Chen pseudo-potential LBE \cite{shanchen} with
hydrophobic/hydrophilic boundaries conditions, as developed in
\cite{prenoi1,prenoi2}. Other models with different boundary conditions and/or
non-ideal interactions have been also used in \cite{kopo}.

\begin{figure}
\centerline{
  \resizebox{0.45\textwidth}{!}{
    \includegraphics{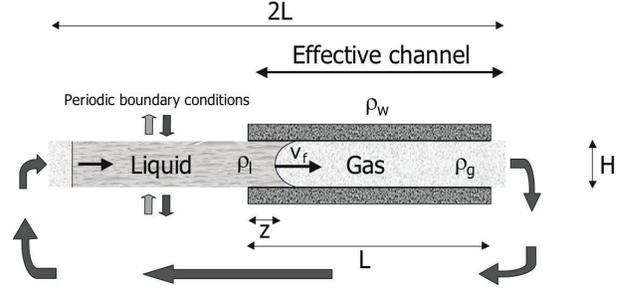}
  }}
   \caption{Geometrical set-up of the numerical LBE. The 2 dimensional
     geometry, with length $2L$ and width $H$, is divided in two
     parts. The left part has top and bottom periodic boundary
     conditions such as to support a perfectly flat gas-liquid
     interface, mimicking a ``infinite reservoir''. In the right
     half, of length $L$, there is the true capillary:  the top and
     bottom boundary conditions are those of a solid wall, with a
     given contact angle $\theta$ \cite{prenoi1}. 
     Periodic boundary conditions are also imposed at the west and east sides. } 
   \label{fig:1}
\end{figure}

The geometry is depicted in
fig. (\ref{fig:1}). The bottom and top surface is coated only in the
right half of the channel, with a boundary condition imposing a given
static contact angle \cite{prenoi1}; in the left half, we impose
periodic boundary conditions at top and bottom surfaces, so as to
realize a flat liquid-gas interface mimicking an ``infinite
reservoir''. Periodic boundary conditions are also imposed at the two
lateral sides, so as to ensure total conservation of mass inside the
system. 

\subsection{LBE algorithm for multi-phase flows}

We start from the usual lattice Boltzmann equation with a single-time
relaxation \cite{Gladrow,Saurobook}: \be\label{eq:LB}
f_{l}(\bm{x}+\bm{c}_{l}\Delta t,t+\Delta
t)-f_l(\bm{x},t)=-\frac{\Delta t}{\tau_B}\left(
  f_{l}(\bm{x},t)-f_{l}^{(eq)}(\rho,\rho {\bm u}) \right) \ee
where $f_l(\bm{x},t)$ is the kinetic probability density function
associated with a mesoscopic velocity $\bm{c}_{l}$, $\tau_B$ is a mean
collision time (with $\Delta t$ a time lapse), $f^{(eq)}_{l}(\rho,\rho
{\bm u})$ the equilibrium distribution, corresponding to the
Maxwellian distribution in the continuum limit.  From the kinetic
distributions we can define macroscopic density and momentum fields as
\cite{Gladrow,Saurobook}: 
\begin{equation} \rho(\x)=\sum_{l} f_{l}(\x); \qquad  \rho
{\bm u}(\x)=\sum_{l}{\bm c}_{l}f_{l}(\x).\end{equation}   For technical details
and numerical simulations we shall refer to the nine-speed,
two-dimensional $2DQ9$ model \cite{Gladrow}. The equilibrium distribution in
the lattice Boltzmann equations is obtained via a low Mach number
expansion of the equilibrium Maxwellian \cite{Gladrow,Saurobook}.  In
order to study non-ideal effects we need to supplement the previous
description with an interparticle forcing. This is done by adding  a
suitable $F_l$ in (\ref{eq:LB}).  In the original  model
\cite{shanchen}, the bulk interparticle interaction is proportional to
a free parameter (the ratio of potential to thermal energy), $G$, entering the equation for the momentum balance:
\be\label{forcing} F_{i}=- G c^{2}_{s}\sum_l w(|{\bm
  c}_{l}|^2) \psi(\x,t) \psi (\x+{\bm c}_l\Delta t,t) {c}^{i}_l \ee
being $w(|{\bm c}_{l}|^2)$ the static weights
for the standard case of 2DQ9 \cite{Gladrow} and
$\psi(\x,t)=\psi(\rho(\x,t))$ the pseudo-potential function which
describes the fluid-fluid interactions triggered by inhomogeneities of
the density profile (see \cite{shanchen,prenoi1,prenoi2} for details).

One may show \cite{prenoi1,prenoi2} that the above pseudo-potential,
leads to a non-ideal pressure tensor given by (upon Taylor expanding the
forcing term): \begin{eqnarray} 
  \label{TENSORE}& P_{ij}&=\left(
    c^{2}_{s}\rho+G\frac{c_{s}^{2}}{2}\psi^{2}+
    G\frac{c^{4}_{s}}{4}|{\bf
      \nabla}\psi|^{2} +G \frac{c^{4}_{s}}{2}\psi \Delta \psi
  \right) \delta_{ij}-  \nonumber \\ &-& \frac{1}{2}{G {{c}}^{4}_{s}}\partial_{i}
  \psi \partial_{j}\psi+ {\cal O} (\partial^4), \end{eqnarray} where
$c_s$ is the sound speed. 
This approach allows the definition of a static contact angle $\theta$, by  means
of a suitable value for the pseudo-potential $\psi_w = \psi(\rho_w)$ \cite{prenoi1},
which can span the range $\theta \in [0^o:180^o]$.
Moreover, it also defines a specific value for the surface tension, 
$\gamma_{lg}$, via the usual integration of the offset between normal and transverse
components of the pressure tensor along the liquid-gas interface
\cite{shanchen,prenoi1,prenoi2}.\\
As to the  boundary conditions on the Boltzmann populations,
the standard bounce-back rule is imposed. One can show that the 
bounce-back rule gives no-slip boundary conditions up to second order
in the Knudsen number in the hydrodynamical  limit of single phase flows \cite{noijfm}. 
In the presence of strong density variations, close to the walls across the 
interface, the velocity parallel to the wall may develop a small slip 
length (of the order of the interface thickness, $\lambda_s \propto
\xi $) which, in turn, allows the interface to move. 
At variance with continuum formulations, in which the slip length is
prescribed a-priori, our mesoscopic model allows slip flow to develop on its
own as a result of the fluid-wall interactions coded in the mesoscopic wall
potential $\psi_w$.
However, it is difficult to control exactly this phenomenon, because
even imposing an exact no-slip boundary condition at the wall \cite{yeomans}, the model 
does develop its own dynamics at the first node away from the wall, thus leading to 
an overall non-zero slip velocity. 
To the purpose of controlling the capillary 
filling, one may reabsorb all these effects within the usual Maxwell slip 
boundary conditions: $ u_{s} = \lambda_s \partial_{n} u$. 
It is easy to show that in presence of a slip velocity, the Poiseuille profile becomes:
\begin{equation}
\label{eq:profile}
u(y) = 6 \frac{ \bar{u} }{H^2} \frac{y(H-y)+\lambda_s H}{1 + 6 \lambda_s/ H}
\end{equation}
where the velocity of the front must be identified with the mean
velocity, $\bar{u} \equiv 1/H \int_0^H u(y) dy = \dot z$.  Therefore, the Washburn law
(\ref{eq:washburn}) becomes:
\begin{equation}
z^2(t) - z^2(0) = A \frac{cos(\theta)}{3} t 
\label{eq:washburn1}
\end{equation}
where 
\begin{equation}
A = 1 + 6 \frac{\lambda_s}{H}  
\end{equation}
accounts for finite-slip slow effects.
Under ordinary conditions, the slip length is of the same order of the
atomistic interaction scale, hence $\xi/H < 10 ^{-3}$, so that slip-effects
can be neglected to all practical purposes.
This situation can drastically change in the presence of superhydrophic effects, although
we shall not be concerned with these problems in the present work.

\subsection{Generalized Washburn equation}

As already remarked many years ago \cite{bousanquet}, the Washburn law
(\ref{eq:washburn}) holds in the limit where inertial forces can be neglected
with respect to the viscous and capillary ones.  
This cannot be true in the early stage of the filling process, where strong
acceleration drives the interface inside the capillary.  
However, putting typical numbers for microfluidic devices
($H \simeq 1 \mu m$, $\gamma \simeq 0.072 N/m$, $\rho_l \simeq  10^{-3} kg/m^3$, $\mu_l
\simeq 10^{-3} Ns/m^2$), it is readily checked that the transient time, $\tau_{diff} =
H \gamma \rho_l/\mu_l^2$, is of the order of a few
nanoseconds, hence completely negligible for most practical purposes.

Another important effect which must be kept in mind when simulating
capillary filling, is the unavoidable ``resistance'' of the gas
occupying the capillary during the liquid invasion.
This is a particular ``sensitive'' issue for LB techniques, because reaching
the typical $1:1000$ density ratio between liquid and gas of experimental 
set up, represents a challenge for most numerical methods, particularly
for multiphase Lattice Boltzmann, typically operating 
in the regime $1:10$ to $1:100$. 
In order to take in to account both effects, inertia and gas dynamics,
one may write down the balance between the total momentum change
inside the capillary and the force (per unit width) acting on the 
liquid$+$gas system (here untilded symbols denote physical units):
\begin{equation}
\frac{d (\dot z M(t))}{dt} = F_{cap}+F_{vis}
\label{eq:momentum}
\end{equation}
where  $M(t)=M_g+M_l$ is
the total mass of liquid and gas inside the capillary at any given
time. The two forces in the right hand side correspond to the capillary force,
$F_{cap} = 2 \gamma cos(\theta)$, and to the viscous drag $F_{vis} =
-2 (\mu_g(L-z) + \mu_lz)\partial_nu(0)$. Following the notation of
fig.(\ref{fig:1}) and the expression for the velocity profile
(\ref{eq:profile}) one obtains the final expression (see also
\cite{napoli} for a similar derivation, without considering the slip
velocity):
\begin{eqnarray}
\label{GWE}
&&(\rho_g(L-z)+\rho_l z)
\ddot z  +  (\rho_l-\rho_g) (\dot z)^2  =  \nonumber \\
 &&2
  \frac{\gamma cos(\theta)}{H} - \frac{12 \dot z}{H^2(1+6 \frac{\lambda_s}{H})
} [(\mu_g (L-z) + \mu_l z)]\label{eq:front2}
\end{eqnarray}
In the above equation for the front dynamics, the terms in the LHS
take into account the fluid inertia. Being proportional either to the
acceleration or to the squared velocity, they become negligible for
long times. Washburn law plus the slip correction (\ref{eq:washburn1}) is therefore correctly recovered 
asymptotically, for $t \rightarrow \infty$, and in the limit when
$\rho_g/\rho_l \rightarrow 0$. The above equation is exact, in the
case where evaporation-condensation effects are negligible,
i.e. when the gas is pushed out of the capillary without offering any
mechanical resistance to the advancing liquid.
This is not the case for most mesoscopic models available in the 
literature \cite{shanchen,yeomans}, based on
a diffusive interface dynamics \cite{jacqmin}. As we shall see, only
when either the limit of thin interface $\xi/H \rightarrow 0$
is reached or when the gas phase density is negligible, $\rho_g /\rho_l \rightarrow 0$, the dynamics given by
(\ref{eq:front2}) is correctly recovered. Otherwise, deviations 
induced by condensation/evaporation effects, are observed, which may result 
in significant departure from the Poiseuille profile inside the gas phase.

\section{Numerical results}

In a previous paper \cite{BIFER}, we pointed out the effects of 
finite-density jumps, interface width and inertia in LB simulations
of capillary filling. In the present work, we keep focus on the details of the
former alone, leaving a more complete discussion to a future and lengthier publication.

\subsection{Finite-density effects} 

\begin{figure}
\centerline{
  \resizebox{0.35\textwidth}{!}{
    \includegraphics{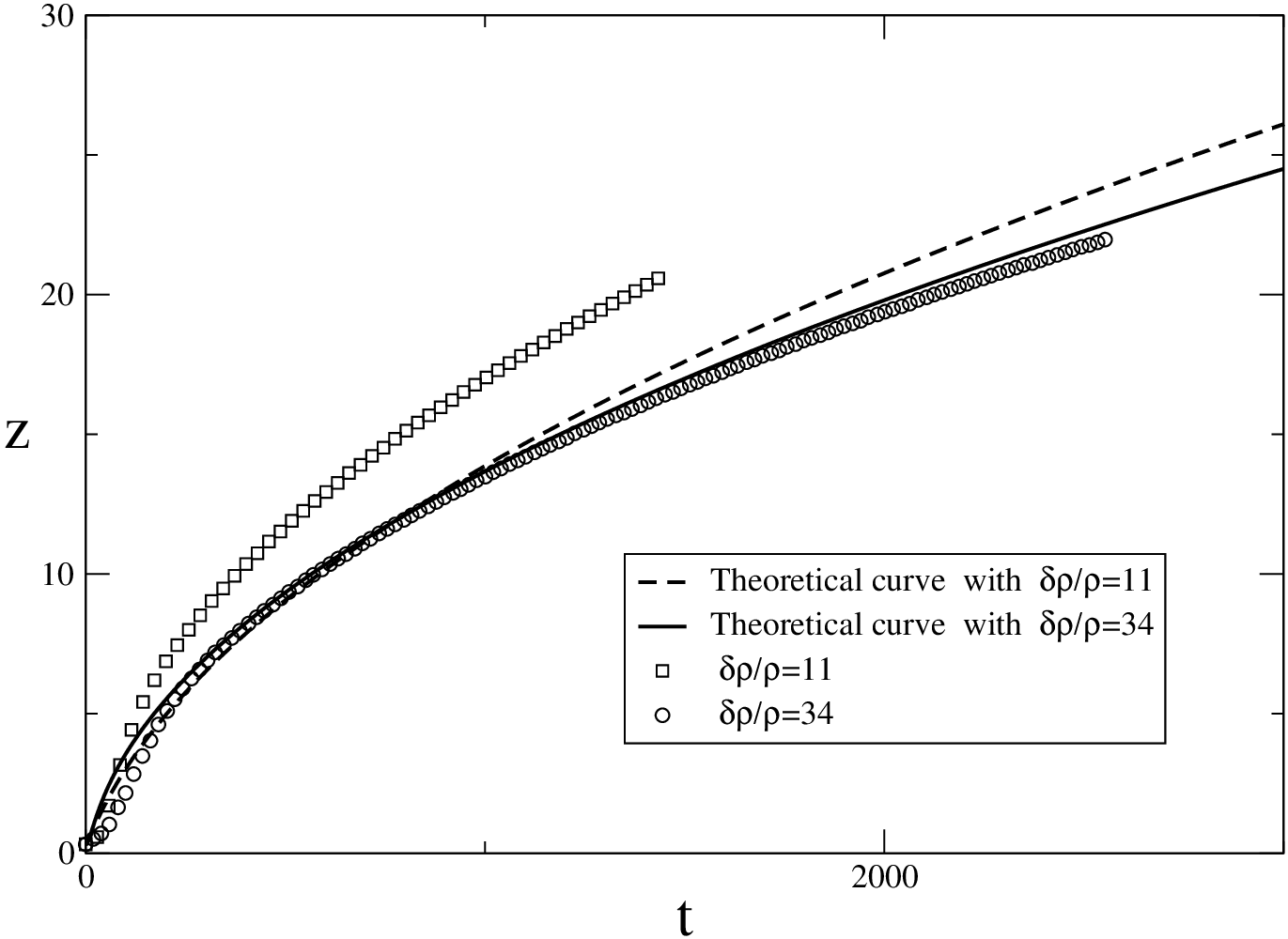}
  }}
\vspace{5mm}
\caption{Evolution of the front coordinate, $z(t)$, for two
values of the density ratio, $\rho_l/\rho_g=34$ and $\rho_l/\rho_g=11$.
For the sake of comparison, the numerical solution of the generalized
Washburn equation with the same value of $\rho_l/\rho_g$ is also reported.
The figure clearly shows that only the case $\rho_l/\rho_g=34$ provides
good agreement between the two.}
\label{FIGRHO}
\end{figure}

In Figure \ref{FIGRHO} we show the front trajectory $z(t)$ for a given
resolution $H=121$ and two different density ratios,
$\rho_l/\rho_g = 11$, ($G=5.0$),
$\rho_l/\rho_g = 34$, ($G=6.0$).
The LB solution is compared against numerical solution of the generalized
Washburn equation (\ref{GWE}).
From this figure, it is clearly appreciated that the Washburn solution 
is quantitatively reproduced only for density ratios above $30$.

\begin{figure}[htb]
  \resizebox{0.45\textwidth}{!}{
    \includegraphics{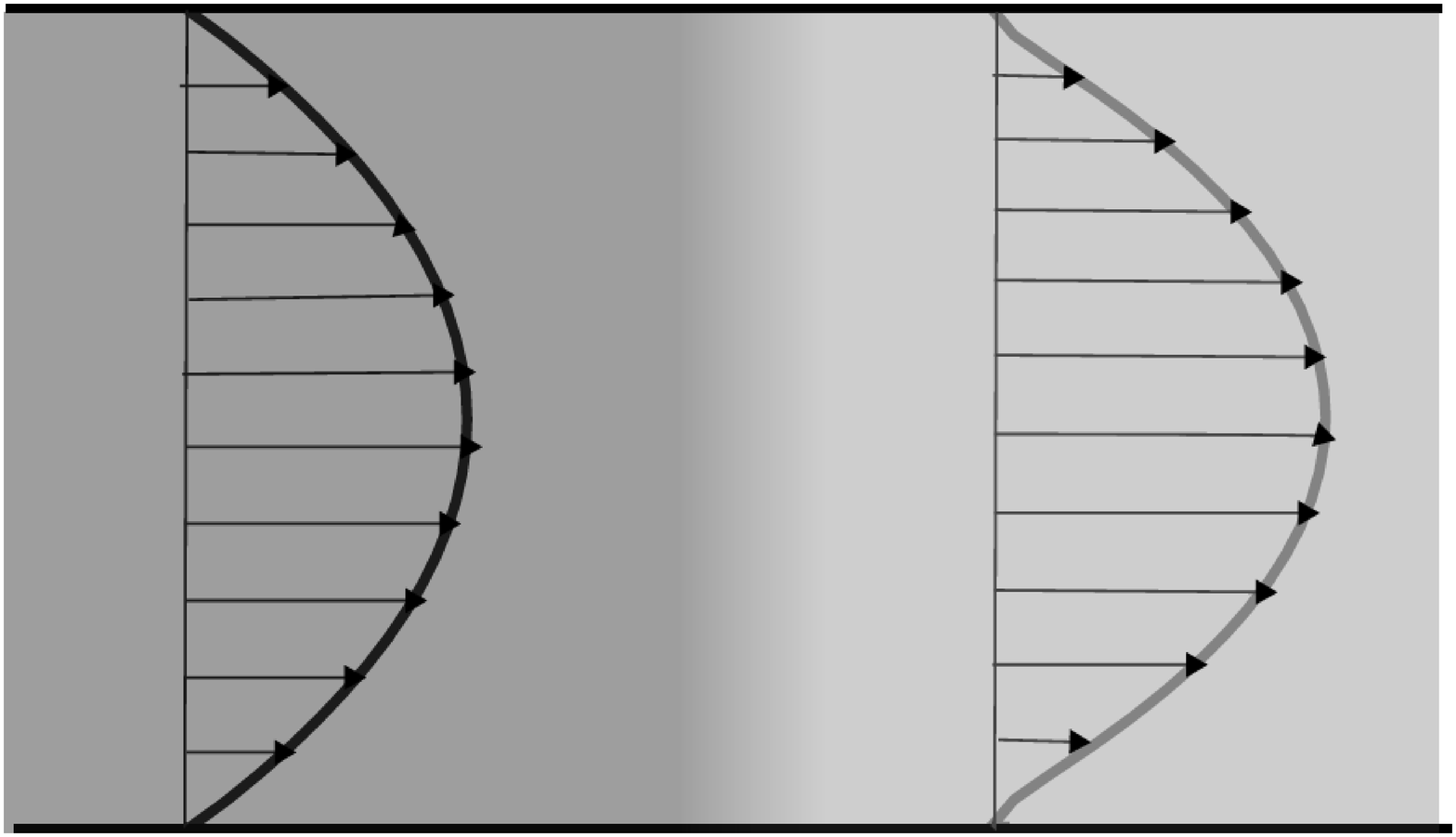}
    \includegraphics{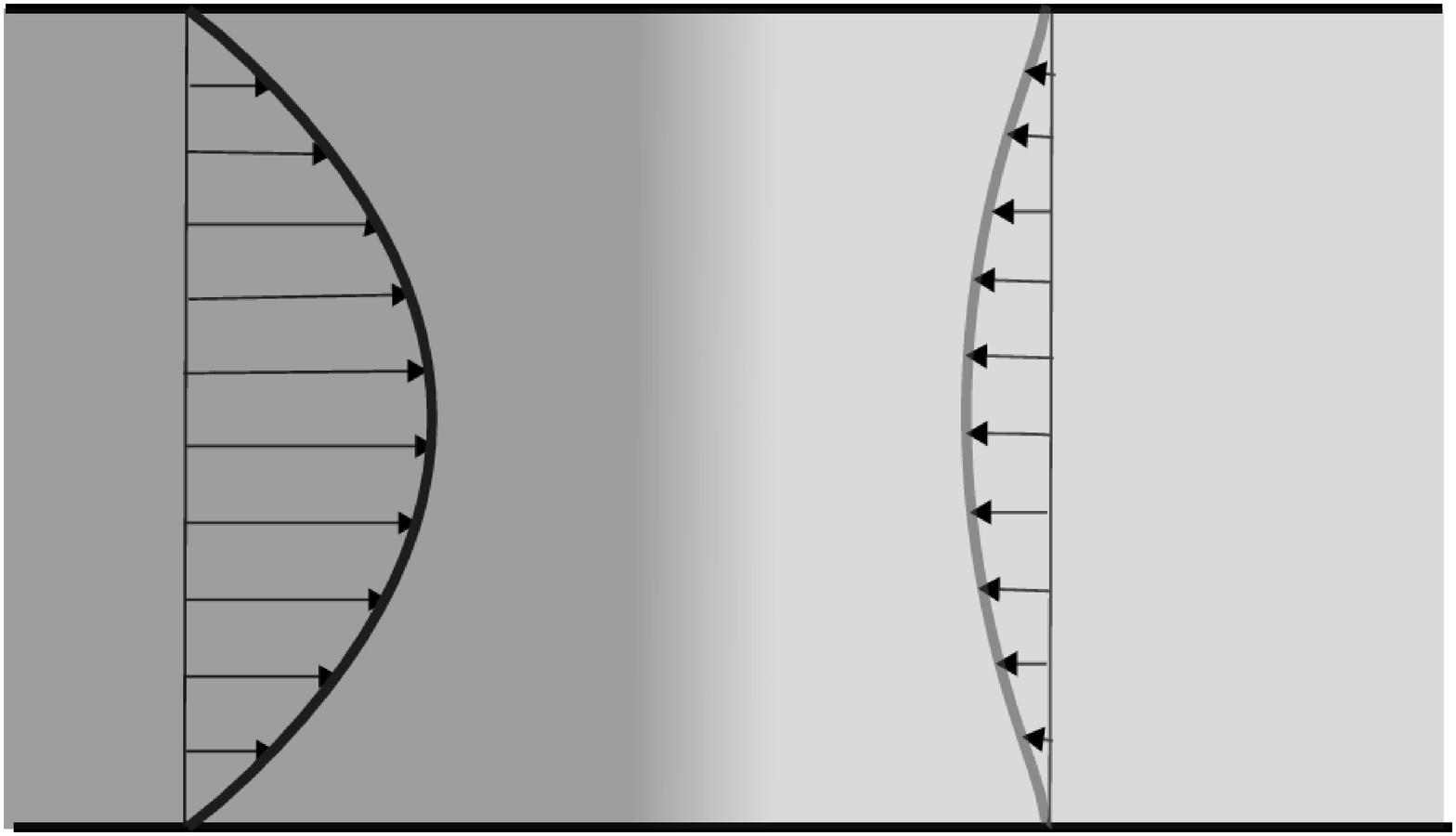}}
\caption{Streamwise velocity field for the case $\rho_l/\rho_g=11$ (left) and
$\rho_l/\rho_g=34$ (right).The dark region marks the dense phase.}
\label{FIGVEL}
\end{figure}

It is instructive to inspect the internal structure of the flow (Fig. \ref{FIGVEL}).
In the left panel, the expected Poiseuille flow is clearly visible for both  
liquid and gas phases. Of course, distortions from such asymptotic behaviour 
take place in the vicinity of the moving interface, but since the vapour phase is light
enough, such distortions are rapidly reabsorbed away from the interface. 
As the vapour phase is made heavier, however, such reabsorption no longer takes place,
and a {\it qualitative} change is observed in the flow pattern, with a clear emergence
of a 'anti-Poiseuille' flow in the gas phase. 
Such anti-Poiseuille flow appears to be a direct
consequence of the fact that, within this parameter regime, the front evolves under the
combined effect of two distinct mechanism: capillary drive {\it and} front-condensation.
The extent to which such condensation holds in real experiments, remains an open issue at
this point, and one to which LB can hopefully contribute useful insights for the future.
Here, we only wish to offer a few remarks.
First, we observe that since front condensation proceeds diffusively, it 
contributes the same $t^{1/2}$ scaling exponent as capillary advection.
This explains why one can recover the correct $1/2$ 
Washburn exponent, and yet miss the right prefactor. 
This is relevant to LB, since it is known
that LB simulations are affected by spurious currents due to
lack of isotropy of the higher-order kinetic moments.
\begin{figure}[htb]
\centerline{
\resizebox{0.45\textwidth}{!}{
\includegraphics{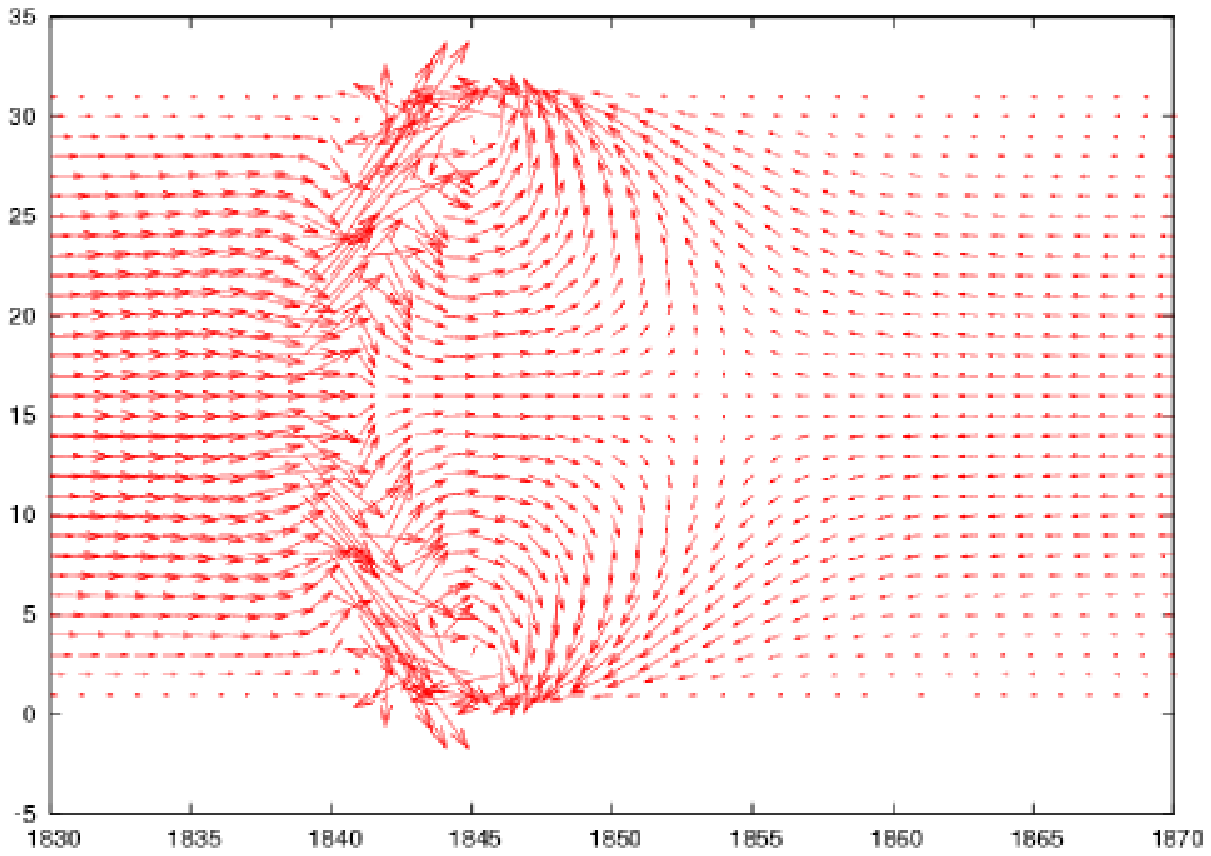}}}
\centerline{
\resizebox{0.45\textwidth}{!}{
\includegraphics{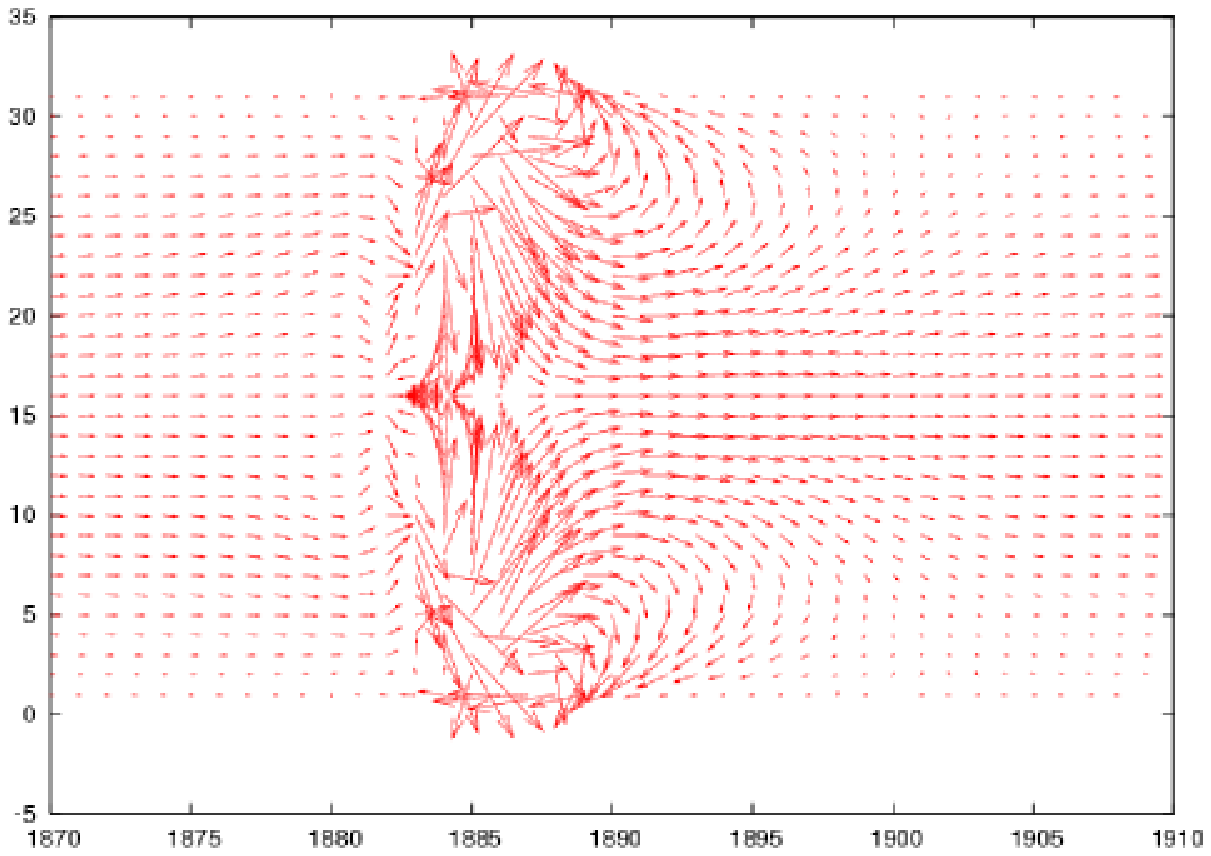}}}
\caption{Two-dimensional velocity pattern for the case $\rho_l/\rho_g=11$ (upper fig.)and $\rho_l/\rho_g=34$ (lower fig.).
The presence of two rolls (vortices) near the solid wall, ahead and past
the front is clearly visible.
Equally visible, is in the upper figure, the inverted flow in the light phase, far ahead
of the front. The far-field velocity ahead of the front of the lower figure, however, recovers the Poiseuille profile.}
\label{FIGVOR}
\end{figure}

One might argue that the rolls shown in figures \ref{FIGVOR}, 
are just a discreteness artifact due to spurious currents. 
We are reasonably confident that this is {\it not} the case, because the same phenomenon
has been observed in independent LB simulations using the momentum-conserving
free-energy formulation, which is known to suffer less severe spurious-current 
effects \cite{YEOMA}.  
Moreover, by running the same case at doubled resolution, the same picture has been obtained, namely
the vortex intensity was found to stay nearly unchanged.
Since it is know that spurious currents decay with grid resolution \cite{prenoi2}, this
provides further confidence that front condensation is a real physical effect
which may occur in experiments near the liquid-gas phase-transition. 
However, since in LB simulations the interface width is far thicker than in
actual experiments, it is possible that the condensation effects 
predicted by the simulation might significantly overestimate the real ones.
It would therefore be desirable to develop a theoretical model to
predict the share of condensation/capillarity as a function of the liquid-gas 
density ratio, possibly by extending the generalized Washburn equation along
the ideas outlined in \cite{POMEAU}. 
Work along these lines, as well as to understand the role of dynamic versus static
contact angle and additional dissipative effects induced by the front 
deformation during propagation, is currently underway.


\section{Conclusions}

The present study shows that Lattice Boltzmann models with
pseudo-potential energy interactions are capable of reproducing the
basic features of capillary filling, as described within a (generalized) Washburn
approximation.  Two conditions for quantitative agreement have been
identified in the recent past: i) a sufficiently high density contrast between the
dense/light phase, $\rho_l/\rho_g > 10$ and a sufficiently thin
interface, $\xi/H < 0.1$.  
In this paper we have focused on the former effect, and shown that at sufficiently
high vapour density (above $1/10$ of the liquid density) front motion proceeds
through a concurrent combination of capillary advection and front condensation.
The latter is sustained by an inverted Poiseulle motion in the light phase, providing
a steady supply of vapour to the front location, where condensation takes place.
Independent LB simulations, based on a different
LB formulation, do confirm the same picture, thereby lending
weight to the hypothesis that at finite vapour density, capillary filling may proceed
through a combination of surface tension drive and condensation.
It is also observed that since both mechanisms feature the same scaling exponent, such
an hypothesis can only be tested through a detailed and quantitative analysis of
the prefactors and not just the exponents. Our data suggest that 
motion by condensation proceeds faster than the capillary one. 
Should this indication receive experimental confirmation, one might 
devise ways of accelerating capillary filling by performing experiments 
near the critical point.

\section{Acknowledgments}

Valuable discussions with B. Andreotti, D. Kwok, D. Palmieri, D. Pisignano, G. Ruocco and J. Yeomans
are kindly acknowledged. 
Work performed under the EC contract NMP3-CT-2006-031980 (INFLUS).
SS wishes to acknowledge support from the Killam Foundation at the University of Calgary.

\end{document}